\documentclass{czjphys}         
\usepackage{epsf}
\usepackage{graphicx}
\newcommand{\gsim}{\lower.7ex\hbox{$\;\stackrel{\textstyle>}{\sim}\;$}}

\begin{document}
\title{Probing Exotic Physics With Cosmic Neutrinos\footnote{This talk was based largely on work done with Jaime Munoz-Alvarez, Luis Anchordoqui, John Beacom, Nicole Bell, Jonathan Feng, Tao Han, Francis Halzen, Dean Morgan, Sandip Pakvasa, Subir Sarkar, Elizabeth Winstanley and Tom Weiler.}}

\authori{Dan Hooper}      
\addressi{Fermi National Accelerator Laboratory, Particle Astrophysics Center, Batavia, IL 60510, USA}

\authorii{}
\addressii{}
\authoriii{}    \addressiii{}
\authoriv{}     \addressiv{}
\authorv{}      \addressv{}
\authorvi{}     \addressvi{}
\headauthor{Dan Hooper}   
\headtitle{Probing Exotic Physics With Cosmic Neutrinos \ldots}
\lastevenhead{Dan Hooper: Probing Exotic Physics With Cosmic Neutrinos 
              ldots}

\refnum{}
\daterec{XXX}    
\issuenumber{0}  \year{2001}
\setcounter{page}{1}
\maketitle
\begin{abstract}
Traditionally, collider experiments have been the primary tool used in searching for particle physics beyond the Standard Model. In this talk, I will discuss alternative approaches for exploring exotic physics scenarios using high energy and ultra-high energy cosmic neutrinos. Such neutrinos can be used to study interactions at energies higher, and over baselines longer, than those accessible to colliders. In this way, neutrino astronomy can provide a window into fundamental physics which is highly complementary to collider techniques. I will discuss the role of neutrino astronomy in fundamental physics, considering the use of such techniques in studying several specific scenarios including low scale gravity models, Standard Model electroweak instanton induced interactions, decaying neutrinos and quantum decoherence.   
\end{abstract}

\section{Introduction: Cosmic Neutrinos as a Probe of Fundamental Physics}     

Particle collisions observed in accelerator experiments have so far been limited to center-of-mass energies below roughly 2 TeV. When the LHC begins operation later this decade, it should be capable of studying hadron collisions with center-of-mass energies up to 14 TeV. A cosmic neutrino with 1 EeV of energy ($10^9$ GeV) scattering with a nucleon at rest, in contrast, has a center-of-mass energy of more than 40 TeV. A 100 EeV neutrino scattering with a nucleon at rest, in turn, corresponds to a center-of-mass energy in excess of 400 TeV. If such interactions could be observed and studied, they could reveal information about particle physics interactions which is inaccessible to current or planned collider technologies.

In addition to this simple arguement of energetics, the observation of cosmic neutrinos provides us with an opportunity to observe particles which have traveled over enormously long baselines. Whereas collider experiments are limited to studying phenomena which occur over very short lengths of time, other techniques can be used to observe properties of particles which emerge over longer lengths of time or, correspondingly, long propagation lengths. For example, long baseline neutrino experiments as well as solar and atmospheric neutrino studies have revealed to us a great deal about neutrino masses and mixing. This behavior would be much more difficult to study effectively in a collider environment. But what about phenomena which might take place over even longer times or propagation lengths? Neutrinos traveling over far greater distances -- kiloparsecs, megaparsecs or gigaparsecs -- are the most useful tool available for studying very long baseline physics.

Of course, these techniques also have their disadvantages. In particular, the high-energy cosmic neutrino flux is far smaller than the flux of particles in an accelerator beam. To use the language of collider physics, this is a limitation of beam luminosity. To compensate for this limitation, neutrino detectors with very large target masses have been developed and/or deployed. These experimental efforts include large volume neutrino telescopes, such as Amanda, Antares, Baikal, Nestor, Rice and IceCube, which use natural water or ice as a detector medium. Ultra-high energy cosmic ray air shower detectors, including the Pierre Auger Observatory (PAO), Agasa, HiRes, Euso, Owl and others can also be used to detect neutrinos. Dedicated ultra-high energy neutrino detectors using radio or acoustic technologies are also being further developed. The Anita experiment -- a radio antenna onboard a balloon flight planned for late in 2006 -- and/or the Pierre Auger Observatory appear likely to observe the first ultra-high energy cosmic neutrino events in the coming years.

In this talk, I will discuss the role of high energy neutrino astronomy in exploring exotic physics scenarios, keeping in mind the advantanges and disadvantages of such techniques in comparison to collider experiments. In particular, I will focus on models with large deviations from the Standard Model charged and neutral current interaction cross sections at very high energies (beyond the range accessible to colliders) and on models with large deviations from the Standard Model (including neutrino masses and mixings) predictions for neutrino phenomenology over very long propagation distances.  

\section{Measuring the Scattering Cross Sections of High Energy Neutrinos}

A measurement of the number of neutrino-induced events observed by a neutrino telescope is effectively a determination of the product of the neutrino scattering cross section (with nucleons or, in the case of the Glashow resonance, with electrons) and the incoming cosmic neutrino flux. Since the high-energy cosmic neutrino spectrum is not {\it a priori} known, an event rate alone cannot be used to determine the neutrino's scattering cross section.

There are ways to get around this conclusion, however. In particular, we can exploit the fact that the Earth is opaque to particles with a scattering cross section with nucleons larger than roughly $\sim 2 \times 10^{-7}$ mb. This cross section corresponds to the Standard Model prediction for neutrinos with an energy of about 100 TeV. At energies above this value, the flux of neutrinos that penetrates through the Earth becomes increasingly suppressed. Neutrino detectors can compare the rate of neutrinos observed from the direction of the Earth (or slightly skimming the Earth in the case ultra-high energy neutrinos) to the rate of neutrinos observed from other directions. In this way, an effective neutrino-nucleon cross section measurement can be attained \cite{[1],[2]}.

For the remainder of this section, I will consider specific models in which the neutrino-nucleon cross section deviates from the Standard Model charged plus neutral current prediction.

\subsection{Low Scale Gravity: Microscopic Black Hole Production}

In models with a fundamental Planck scale not far above the electroweak scale, gravitational interactions can become quite strong at energies accessible to high energy neutrino detectors. Models with large, or warped extra dimensions are possible realizations of such a scenario. 

The growth of the neutrino-nucleon cross section can be the result of several types of processes, including Kalzua-Klein graviton exchange, string resonances or microscopic black hole production. Of these, black hole production is expected to dominate in the high energy limit.

A naive estimate of the neutrino-nucleon cross section for producing a microscopic black hole follows from the simple geometric description
\begin{equation}
\sigma_{\rm{BH}}(E_{\nu}) \sim \pi r^2_{\rm{sch}}(E_{\rm{CM}}),
\end{equation}
where $r_{\rm{sch}}$ is the Schwartchild radius of a black hole with a mass $E_{\rm{CM}}$. In $4+n$ dimensions, this is given by~\cite{[3]}
\begin{equation}
r_{\rm{sch}}(M_{\rm{BH}}) = \frac{1}{M_D} \bigg[\frac{M_{\rm{BH}}}{M_D} \bigg]^{1/(1+n)} \bigg[\frac{2^n \pi^{(n-3)/2}\,\Gamma(\frac{3+n}{2})}{2+n}\bigg]^{1/(1+n)}.
\end{equation}
This estimate does not take into account that a large fraction of the total energy will likely be radiated away in the form of gravitational waves, thus reducing the energy threshold for black hole production~\cite{[4]}. If the energy that becomes trapped in the form of a microscopic black hole is $M_{\rm{BH}}= y \sqrt{\hat{s}}$, where $y$ is a function of the impact parameter, then the neutrino-nucleon cross section for black hole production can be written as \cite{[5]}
\begin{equation}
\sigma_{\rm{BH}}(E_{\nu}, x_{\rm{min}}, n, M_D) = \int^1_0 dz 2z \int^1_{(x_{\rm{min}} M_D)^2/y^2 s} dx F(n) \pi r_{\rm{sch}}^2(\sqrt{\hat{s}},z) \, \sum_i f_i(x,Q),
\end{equation}
where $z$ is the impact parameter divided by the maximum impact parameter, the maximum impact parameter is $\sqrt{F(n)} r_{\rm{sch}}(\sqrt{\hat{s}},z)$, $F(n)$ is the form factor and $f_i(x,Q)$ are the parton distribution fucntions. The black hole production cross section through neutrino-nucleon scattering is shown in figure~\ref{bhcross}.

Once generated, microscopic black holes evaporate via Hawking radiation with a (rest frame) lifetime on the order of TeV$^{-1} \sim 10^{-27}$ s. The number of particles produced in such a process can vary from a few for PeV neutrinos to tens or hundreds at much higher energies. The types of particles radiated are distributed according to their degrees of freedom. On average, about 75\% of a black hole's energy goes into a hadronic shower, while about 4\% go into electromagnetic radiation. Only about 2.5\% of the particles radiated from an evaporating black hole are tau leptons. 
\bfg[t]
\bc  
\mbox{
\includegraphics[width=3.0in,angle=0]{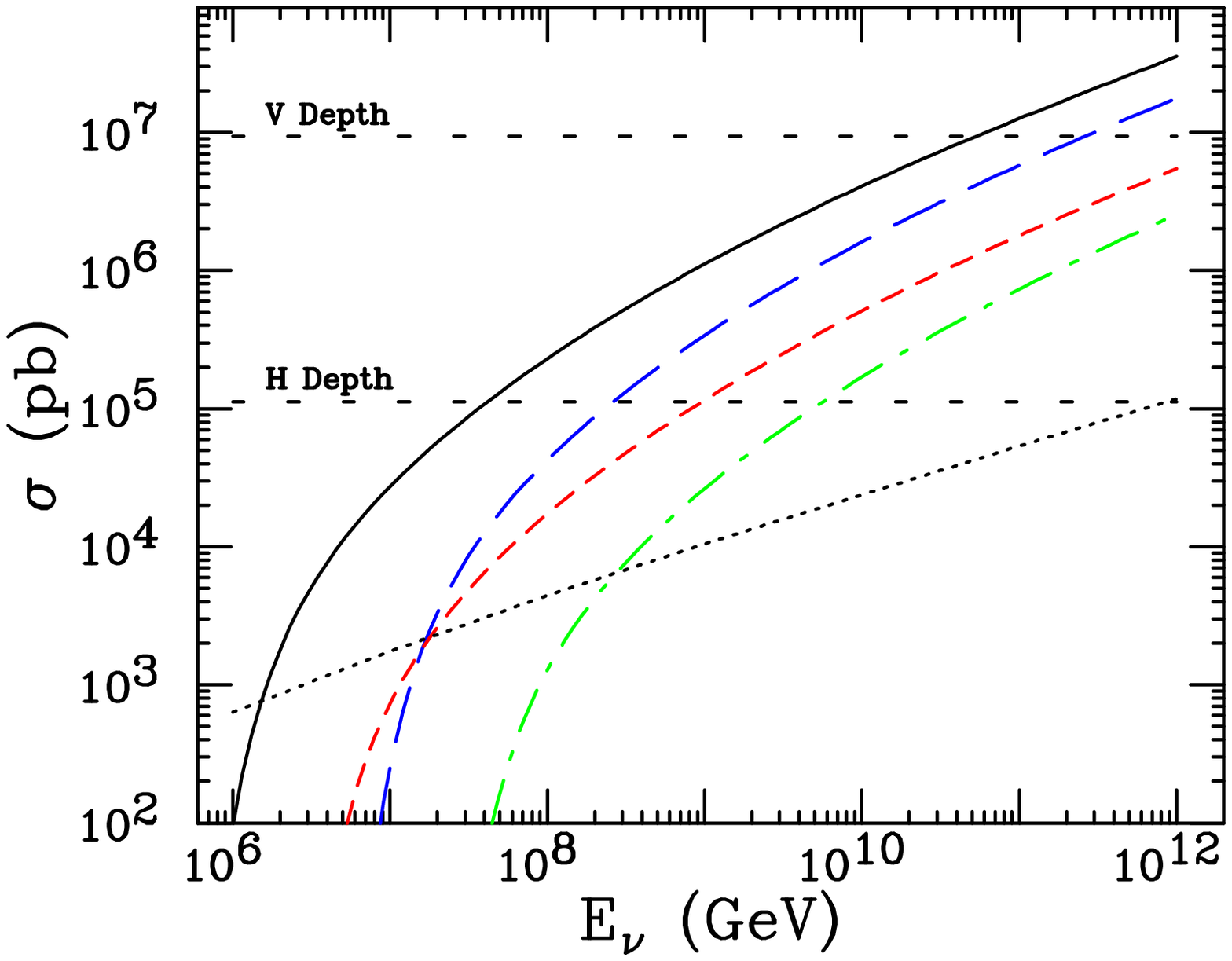} 
}
\ec                         
\vspace{-8mm}
\caption{The cross section for BH production in neutrino nucleon 
  collisions for $n = 6$ extra dimensions, assuming $M_{D} = 1$~TeV 
  and $M_{\rm BH, min} = M_{D}.$ Energy losses by gravitational radiation   
  have been included. The SM $\nu N$ cross section is indicated by the dotted 
  line. For comparison the typical  $pp$ cross section is shown, as well as  
  the cross section required for triggering vertical and horizontal  
  atmospheric showers. The cross section for absorption by the Earth is also  
  shown. Figure originally appeared in Ref.~\cite{[6]}.} 
\label{bhcross} 
\efg 

As a result of the large enhancement in the neutrino-nucleon cross section in these models, the ratio of quasi-horizontal air showers to Earth-skimming tau neutrino induced air showers at Auger can be strongly modified from the Standard Model prediction~\cite{[8]}.  In figure~\ref{bhauger}, the spectrum of neutrino-induced events at Auger is shown including the effects of black hole production~\cite{[9]}. In the left and right frames are the quasi-horizontal and Earth-skimming rates, respectively.  As expected, the quasi-horizontal rate can be enhanced considerably, while the Earth-skimming rate is suppressed. In table~\ref{bhtable} the ratio of these event classes is given. 

\bfg[t]
\bc  
\mbox{
\includegraphics[width=2.0in,angle=90]{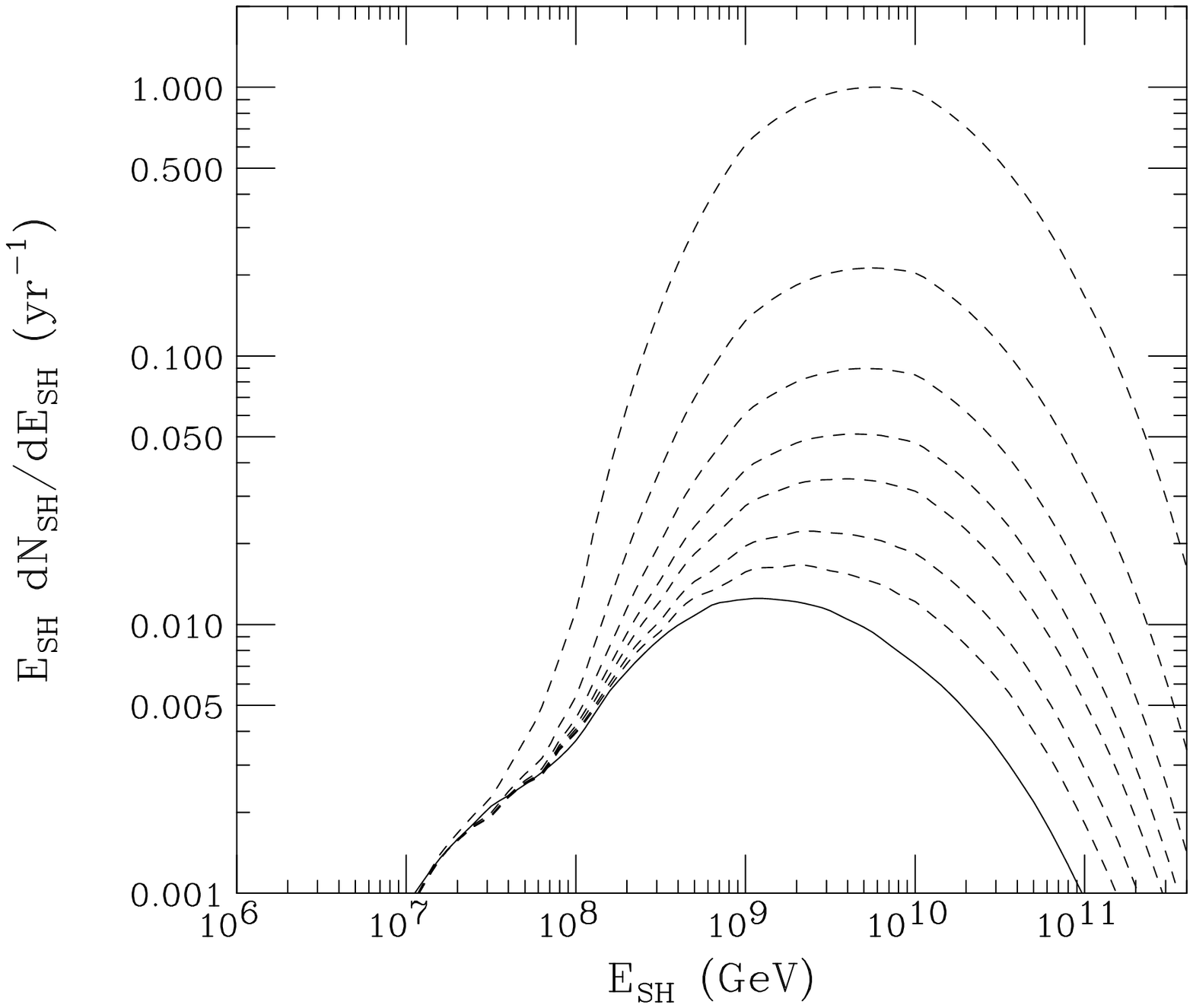} 
\includegraphics[width=2.0in,angle=90]{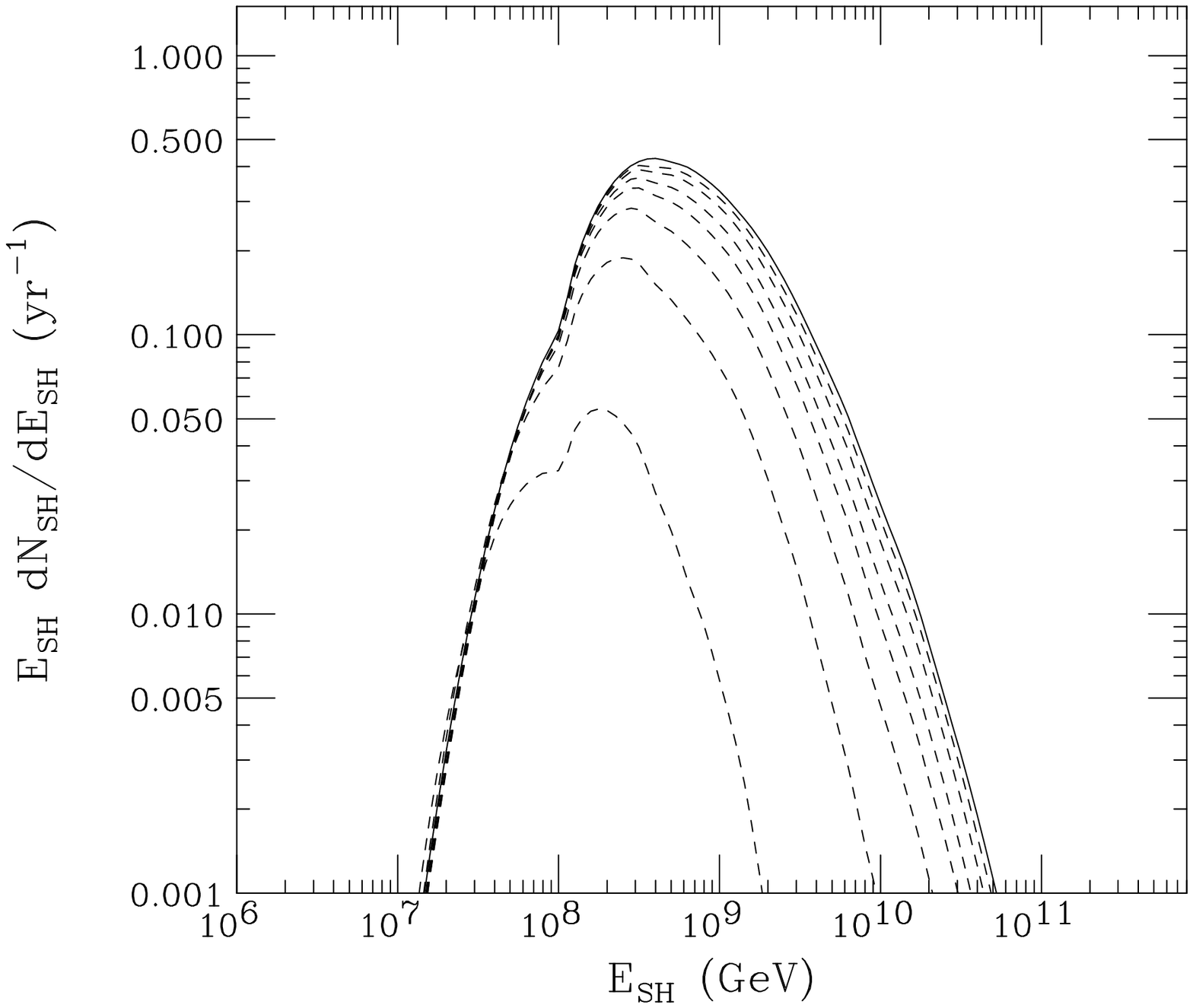} 
}
\ec                         
\vspace{-8mm}
\caption{The spectrum of neutrino induced events predicted at Auger in TeV scale gravity models with microscopic black hole production. The left and right frames correspond, respectively, to quasi-horizontal, deeply penetrating showers and Earth-skimming tau neutrino induced showers. The solid line represents the SM prediction, while the dashed lines include black hole production with fundamental Planck scales of 1, 2, 3, 4, 5, 7 and 10 TeV (from top to bottom in the left frame, and from bottom to top in the right frame).  In each case, $M_{\rm BH, min} = 3 M_{D}$. The cosmogenic neutrino flux of Ref.~\cite{[7]} has been used.  Figure originally appeared in Ref.~\cite{[9]}.} 
\label{bhauger} 
\efg 

\bt[t] 
\caption{Event rates per year at Auger for quasi-horizontal showers 
and Earth-skimming tau neutrino induced showers in low scale gravity models with a range values for the fundamental Planck scale~\cite{[9]}. In each case, $M_{\rm BH, min} = 3 M_{D}$. These rates were calculated using the cosmogenic neutrino flux of Ref.~\cite{[7]}.}
\label{bhtable}
\vspace{2mm}
\small
\bc
\btu{|c|c|c|c|}  
\hline 
\raisebox{0mm}[4mm][2mm] {$\sigma_{\nu N}$} & Quasi-Horizontal &
Earth-Skimming $\nu_{\tau}$ & Ratio  \\ 
\hline \hline 
Standard Model & 0.067 & 1.30 & 0.05  \\ 
\hline 
$M_{D}=$1 TeV & 4.400 &0.13 & 36.00  \\ 
\hline 
$M_{D}=$2 TeV & 0.950 & 0.48 & 2.00  \\
\hline 
$M_{D}=$3 TeV & 0.420 & 0.77 & 0.54 \\ 
\hline 
$M_{D}=$4 TeV & 0.250 & 0.96 & 0.26  \\ 
\hline 
$M_{D}=$5 TeV & 0.180 & 1.10 & 0.16  \\ 
\hline 
$M_{D}=$7 TeV & 0.120 & 1.20 & 0.10  \\ 
\hline 
$M_{D}=$10 TeV & 0.089 & 1.3 & 0.08  \\ 
\hline \hline
\etu 
\ec
\et

In addition to measuring the enhanced cross section associated with microscopic black hole production, it may be possible to identify the spectrum of products generated in such interactions associated with Hawking radiation. This may be somewhat difficult to accomplish at Auger, as only tau induced events can be identified, and with a considerably suppressed rate. At IceCube, however, events can be identified as muon tracks, electromagnetic and hadronic showers, or as "tau unique" events, such as double bangs or lollipops~\cite{[10]}. By comparing the ratios of these three classes of events, an experiment such as IceCube could play an important role in identifying the characteristics of the spectrum of Hawking radiation.

\subsection{Standard Model Electroweak Instanton Induced Interactions}

Another scenario in which neutrino-nucleon interactions may deviate substantially from the SM charged plus neutral current prediction at very high energies is if non-perturbative electroweak interactions play an important role. Transitions between degenerate vacuua may be possible without suppression at energies above the Sphaleron mass scale. In such a model, it is possible that neutrino-nucleon scattering will suddenly become dramatically enhanced at ultra-high energies, even exceeding the mb level. In figure~\ref{andreas}, the allowed regions for the neutrino-nucleon cross section are shown in this scenario, based on an interpolation between the electroweak and QCD-like behavior, consistent with existing data. For discussion of related phenomenological issues, see Ref.~\cite{[12]}.

\bfg[t]
\bc  
\mbox{
\includegraphics[width=3.0in,angle=0]{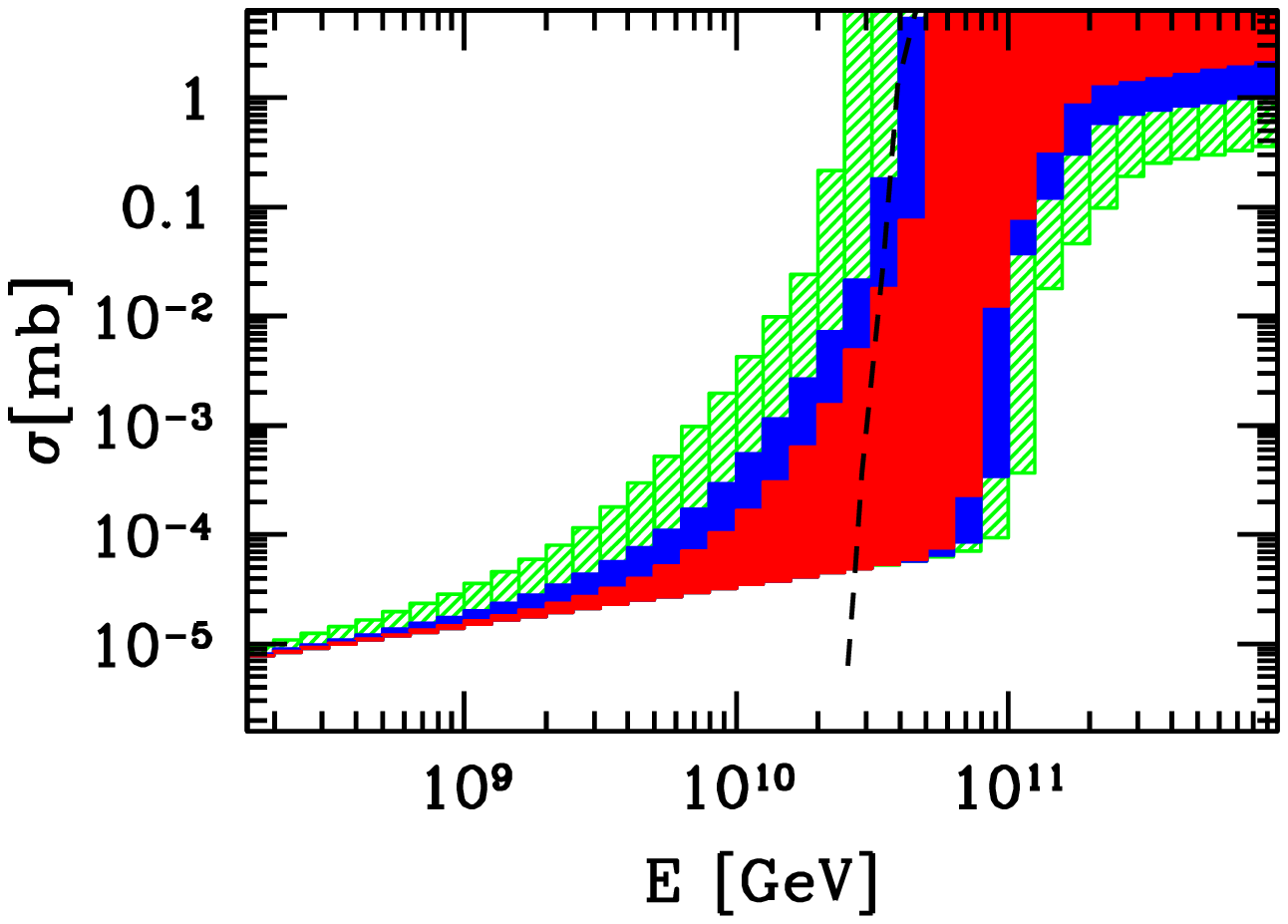} 
}
\ec                         
\vspace{-8mm}
\caption{The allowed 90\%, 95\%, and 99\% confidence level regions for interpolation between the electroweak and QCD-like neutrino-nucleon cross-section
consistent with existing data due to the effects of SM electroweak instanton induced interactions. Figure originally appeared in Ref.~\cite{[11]}.} 
\label{andreas} 
\efg 
\bfg[t]
\bc  
\mbox{
\includegraphics[width=2.0in,angle=90]{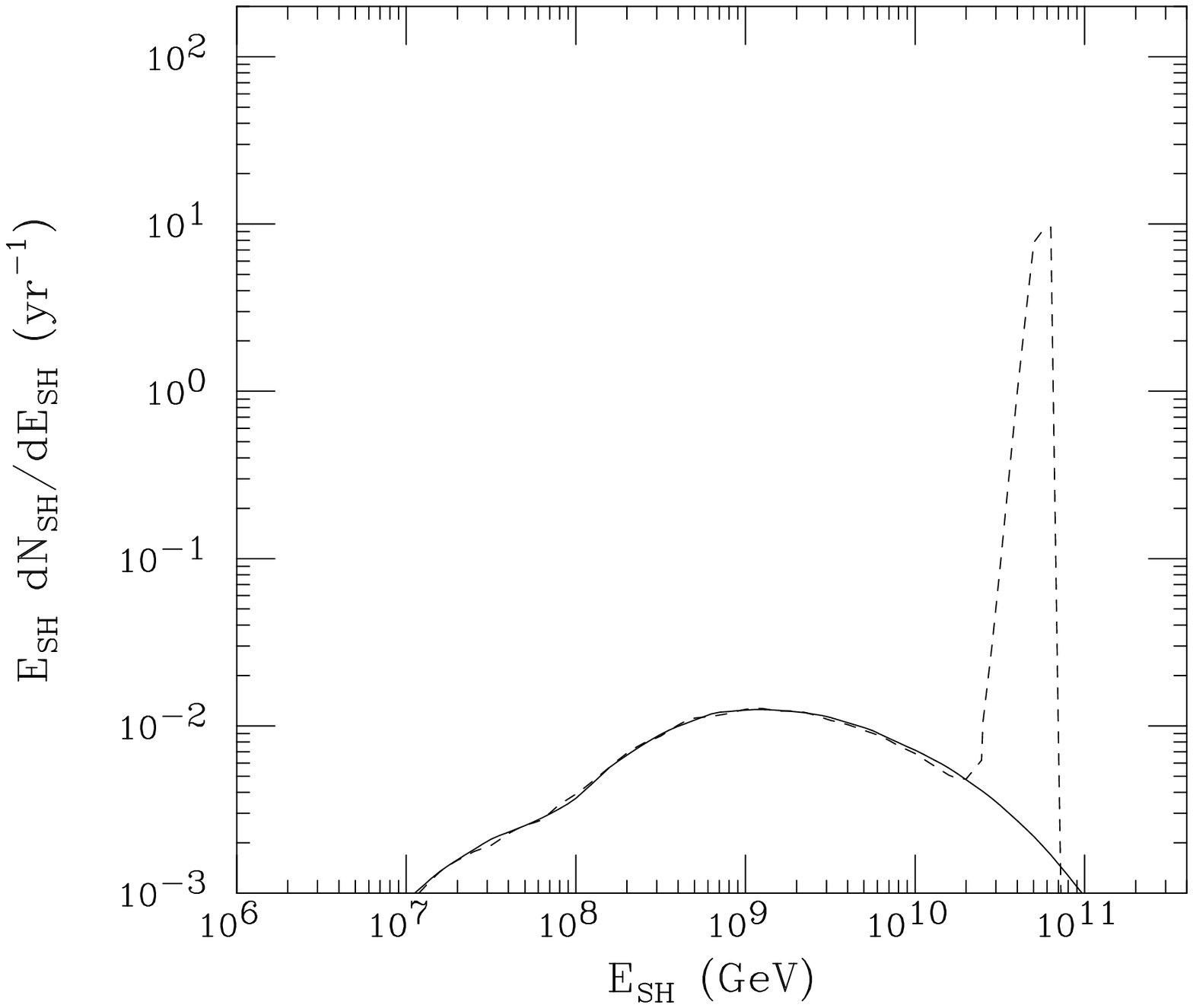} 
\includegraphics[width=2.0in,angle=90]{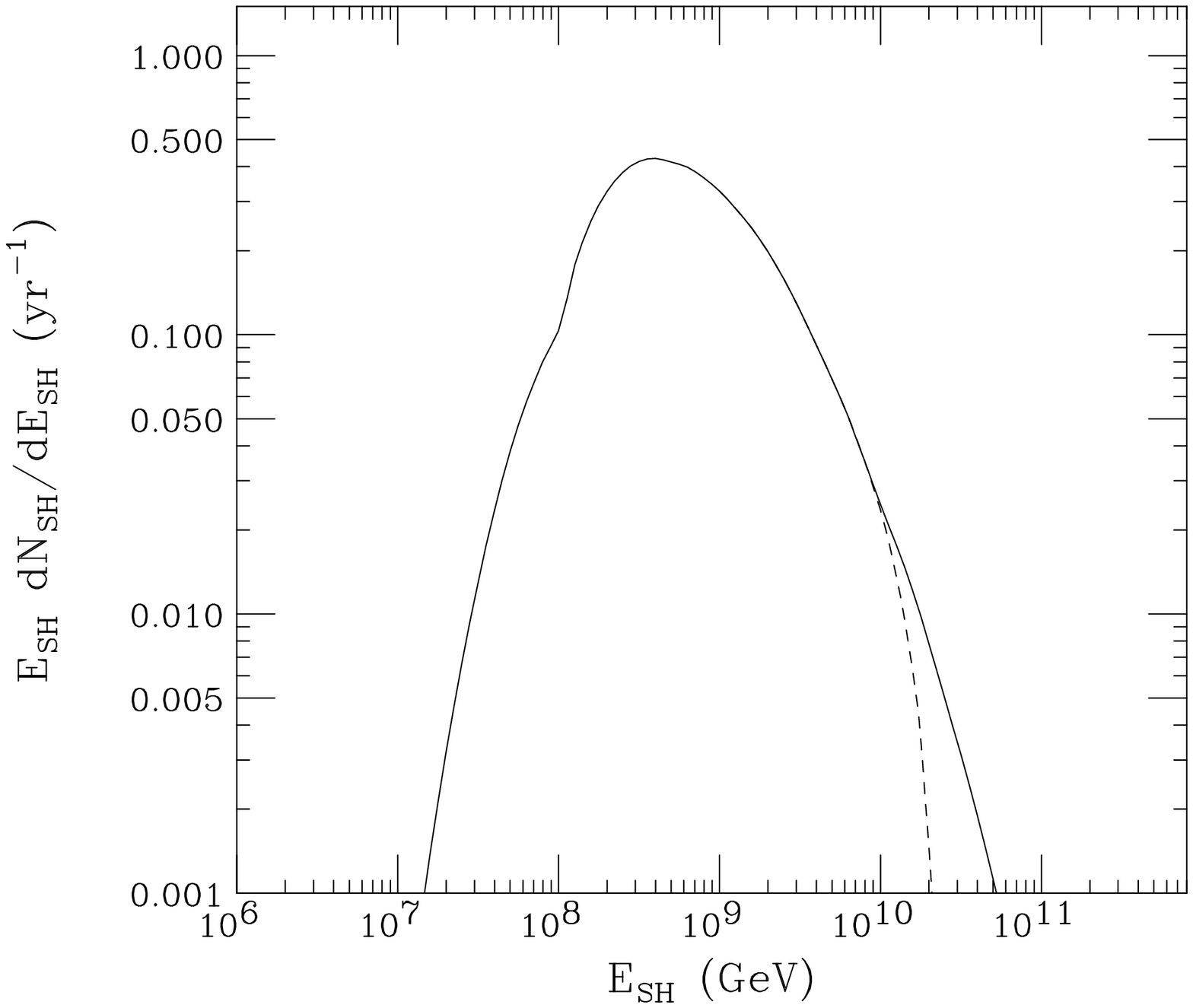} 
}
\ec                         
\vspace{-8mm}
\caption{The spectrum of neutrino induced events predicted at Auger including the effects of non-perturbative electroweak interactions. The left and right frames correspond, respectively, to quasi-horizontal, deeply penetrating showers and Earth-skimming tau neutrino induced showers. The solid line represents the SM charged plus neutral current prediction, while the dashes lines include non-perturbative physics. The cosmogenic neutrino flux of Ref.~\cite{[7]} has been used.  Figure originally appeared in Ref.~\cite{[9]}.} 
\label{insauger} 
\efg 

Such enormous deviations from the charged plus neutral current prediction are ideally suited for an experiment such as Auger. The spectrum of quasi-horizontal, deeply penetrating showers and Earth-skimming tau induced showers in this model are shown in figure~\ref{insauger} \cite{[9]}. Note that the event rate grows in the left frame only over a short range of energies before becoming suppressed. This is the result of atmospheric attenuation which quickly sets in with such rapid cross section growth. The prospects at IceCube in this scenario are potentially interesting, but challenging due to the small event rate anticipated at ultra-high energies \cite{[13]}.

\section{Very Long Baseline Neutrino Measurements}

In addition to performing neutrino-nucleon cross section measurements, high-energy cosmic neutrinos may be used to learn about the properties of neutrinos exhibited only over very long timescales or propagation lengths. Such tests are carried out by measuring the ratios of neutrino flavors present in the cosmic neutrino spectrum. Deviations in these ratios from the SM expectations can be an indicator of new physics.

Cosmic neutrinos are thought to be primarily generated in processes involving the decay of charged pions. Such decays, which proceed as $\pi^+ \rightarrow \mu^+ \nu_{\mu} \rightarrow e^+ \nu_e \bar{\nu}_{\mu} \nu_{\mu}$, generate neutrinos with flavors in the ratio of $\nu_e:\nu_{\mu}:\nu_{\tau} = 1:2:0$. After the effects of oscillations are included, this ratio evolves very closely to $\nu_e:\nu_{\mu}:\nu_{\tau} = 1:1:1$. This is a fairly robust prediction, although it may be altered margainly due to the effects of energy losses in the source before the pion or muon decays are completed \cite{[14]}.

A experimental confirmation of the ratio $\nu_e:\nu_{\mu}:\nu_{\tau} = 1:1:1$ among cosmic neutrinos would provide a strong constraint on a range of exotic physics scenarios. I will now turn to two examples of such physics: decaying neutrinos and quantum decoherence.

\subsection{Neutrino Decay}

Neutrinos are known to be remarkably stable against decays through channels such as $\nu_i \to \nu_j \gamma$ or $\nu \to \nu \nu \overline \nu$. Furthermore, at the tree level, they are entirely stable to electroweak decays. Other possibilities leading to neutrino instability are less well constrained, however. In particular, processes such as $\nu_i \rightarrow \nu_j  X$ or $\nu_i \rightarrow
\overline{\nu}_j X$, are possible, where $\nu_{i,j}$ denote mass
eigenstates and $X$ denotes a Majoron or other such light or massless state. Current constraints on such possibilities are
rather weak, and come from Solar neutrino data, which can be used to set the bound, $\tau/m \gsim 10^{-4}$ s/eV~\cite{[15]}. By
studying cosmic neutrinos which have travelled over far longer
baselines, however, neutrino telescopes can be far more sensitive to neutrino instability if an effective flavor ratio measurement can be
made.

The ratios of flavors observed in the cosmic neutrino spectrum depends
on whether any species of neutrinos have decayed and, if so, through what decay
channel. In the simple situation where all heavy neutrino species
decay into the lightest mass eigenstate (or into non-interacting
states, such as a sterile neutrino), we would expect to observe at
Earth the flavor ratio
\begin{equation}
\phi_{\nu_e}:\phi_{\nu_{\mu}}:\phi_{\nu_{\tau}} = 
\cos^2\theta_{\odot} : \frac{1}{2} \sin^2 \theta_{\odot} 
:\frac{1}{2} \sin^2 \theta_{\odot} \approx 6:1:1, 
\end{equation}
where $\theta_{\odot}$ is the solar neutrino mixing angle and we have
assumed the normal hierarchy as well as $U_{e3} =0$.  This result is
independent of the flavor ratio at source.  However, for the case of
an inverted hierarchy, the predicted flavor ratio at Earth is
\begin{equation}
\phi_{\nu_e}:\phi_{\nu_{\mu}}:\phi_{\nu_{\tau}} = 
U^2_{e3}:U^2_{\mu 3}:U^2_{\tau 3} \approx 0:1:1,
\end{equation}
where $U_{\alpha_i}$ is the neutrino mixing matrix and we have taken
the atmospheric mixing angle to be maximal.  These results are in
striking contrast to the expectation for stable neutrinos.

\bt[t]
\caption{The neutrino flavor ratios predicted for a variety of 
neutrino decay models with decay mode as indicated~\cite{[16]}.}
\label{decaytable}
\small
\bc
\btu{|c|c|c|} 
\hline 
\raisebox{0mm}[4mm][2mm] Decaying Mass Eigenstates & Decay Products &
 $\phi_{\nu_e}:\phi_{\nu_{\mu}}:\phi_{\nu_{\tau}}$ \\
 \hline \hline
 $\nu_3$, $\nu_3$ & Irrelevant & 6:1:1  \\
\hline
 $\nu_3$ & Invisible & 2:1:1   \\
 \hline
 $\nu_3$ & $\nu_2$ & 1.4\,--1.6:1:1   \\
 \hline
 $\nu_3$ & $\nu_1$ & 2.4\,--2.8:1:1   \\
 \hline
 $\nu_3$ & 50\%\,$\nu_1$, 50\%\,$\nu_2$ & 2:1:1   \\
 \hline \hline
 \etu
\ec
\et
\bfg[t]
\bc  
\mbox{
\includegraphics[width=2.6in,angle=90]{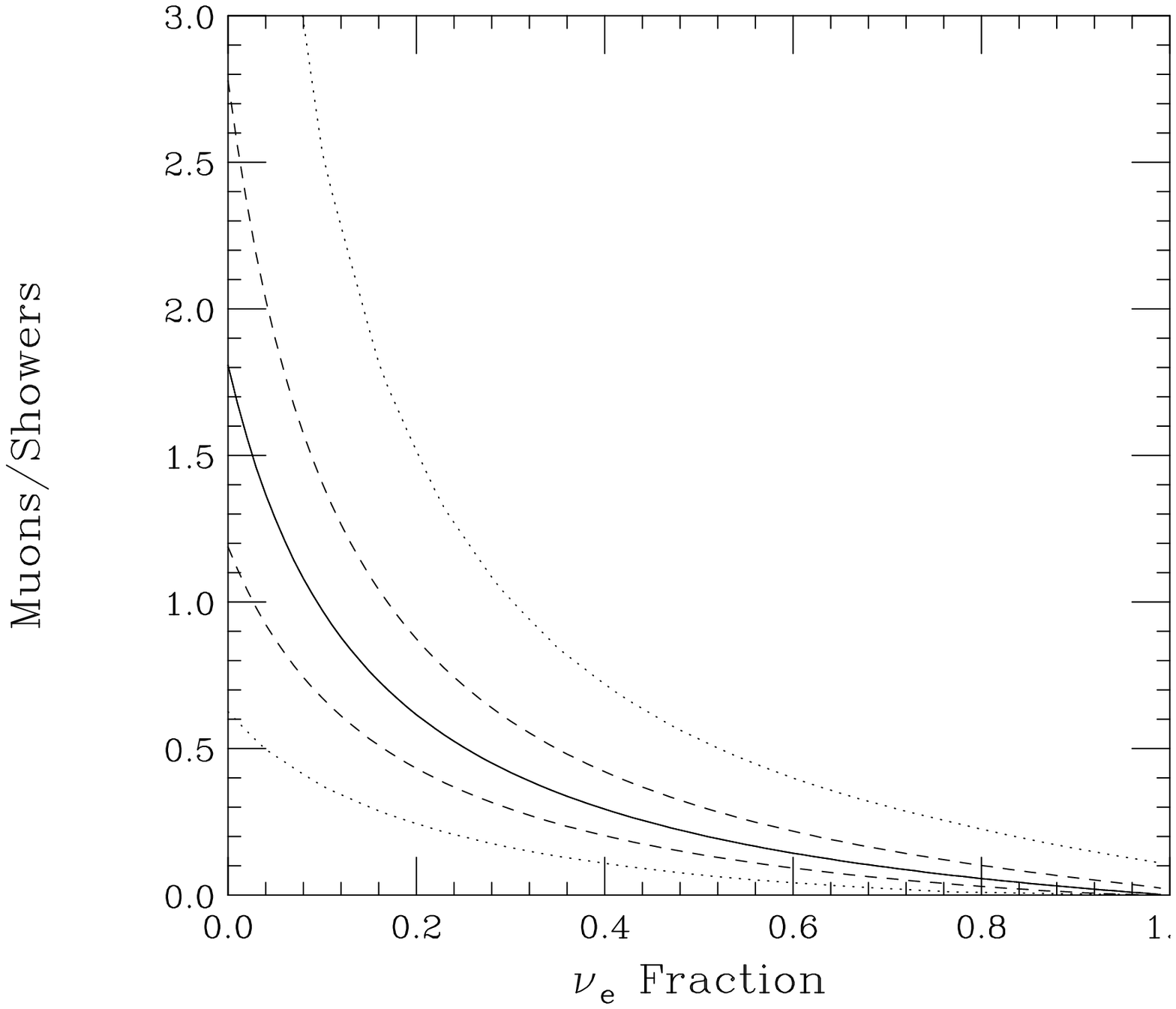} 
}
\ec                         
\vspace{-8mm}
\caption{The ability of a kilometer-scale high-energy neutrino telescope such as IceCube to measure the flavor ratios of the cosmic neutrino spectrum. The quantity along the y-axis denotes the ratio of muon tracks (with a threshold of $E_{\mu} > 1$ TeV imposed) to shower events observed. The solid line represents the true relationship between this observable ratio and the $\nu_e$ content of the neutrino flux. The dashed and dotted lines represent the measured ranges (at the 68\% confidence level) for a spectrum of $E^2_{\nu_{\mu}} dN_{\nu_{\mu}}/dE_{\nu_{\mu}} = 10^{-7}$ GeV cm$^{-2}$ s$^{-1}$ and $2 \times 10^{-8}$ GeV cm$^{-2}$ s$^{-1}$, respectively. Figure originally appeared in Ref.~\cite{[17]}.} 
\label{decaymeasure} 
\efg 

These two cases represent the most extreme deviations from the usual
phenomenology and are robust in that they do not depend on the flavor
composition at source. A variety of other possibilities have also been considered~\cite{[16]}.  In Table~\ref{decaytable} we list some of these possibilities assuming
the usual mass hierarchy and source flavor ratios as for pion decay.

Neutrino telescopes do not directly measure the flavor ratios of the cosmic neutrino spectrum, but instead detect classes of events. The ratio of these event types can, in principle, be used to infer the corresponding flavor ratios. At IceCube, these event classes consist of muon tracks, electromagnetic or hadronic showers, and "tau unique" events. Auger, on the other hand, only identifies quasi-horizontal showers and Earth-skimming tau neutrino induced events. 

In figure~\ref{decaymeasure} the ability of a kilometer-scale neutrino telescope, such as IceCube, to determine the flavor ratios of the cosmic neutrino flux is shown. By measuring the ratio of muon tracks to shower events observed, a rough determination of these ratios can be made, given a sizable cosmic neutrino flux, observation time, or both.

\subsection{Quantum Decoherence}

Finally, I will discuss the prospects for identifying the effects of quantum decoherence using cosmic neutrinos. Quantum decoherence is a feature expected in some quantum gravity frameworks, and represents processes involving the loss of quantum information. In particular, the flavor information contained in a flux of neutrinos may be erased over time. Cosmic neutrinos represent the longest baseline, and therefore potentially the most sensitive probe of these effects.

The generic predition of quantum decoherence is the transition toward neutrino flavors with the ratio $\nu_e:\nu_{\mu}:\nu_{\tau} = 1:1:1$, regardless of their initial flavor composition. This is, unfortunately, very similar to the prediction for neutrinos generated through pion decay. Therefore another source of cosmic neutrinos is needed to search for the effects of quantum decoherence.

In addition to pion decays, cosmic neutrinos may also be generated through the decay of neutrons, producing uniquely electron anti-neutrinos. Such neutrons might come from the photo-dinintegration of ultra-high energy nuclei, for example. If such a source could be identified, and no quantum decoherence effects are present, standard oscillations would modifiy the (anti-)neutrinos' flavor ratio as $\nu_e:\nu_{\mu}:\nu_{\tau} = 1:0:0 \rightarrow 0.56:0.24:0.20$.  If such a deviation from $1:1:1$ could be measured, it would place very strong constraints on quantum decoherence \cite{[18]}.

Galactic cosmic ray accelerators may be particularly useful in this application. EeV neutrons have a decay length on the order of 10 kiloparsecs. Being electrically neutral, EeV neutrons produced at accelerators within the milky way are not deflected by magnetic fields and will appear as point sources to UHE cosmic ray experiments. There is some evidence in the UHECR data for the existence of EeV cosmic ray point sources~\cite{[19]}, of which the Cygnus region is perhaps the most promising.

Although some of the neutrons generated in galactic sources will reach Earth intact, others will decay producing potentially observable numbers of anti-neutrinos~\cite{[20]}. If future observations confirm the existence of an EeV neutron source in Cygnus region, this source would represent the laboratory with the greatest prospects to constrain (or discover) the effects of quantum decoherence \cite{[18],[21]}.

\section{Conclusions}

Collider experiments have been the single greatest tool for exploring many of the most important questions in particle physics. As time goes on, however, we are gradually witnessing these experiments becoming increasingly technologically challenging and costly to construct. I have little doubt that collider physics will remain a viable and exciting way of exploring particle physics for decades to come. That being said, there will also likely be a point in the future at which other experimental approaches may become more practical or cost effective. With this eventuality in mind, it is important to consider now how we can explore fundamental physics using all of the tools at our disposal -- in particular, those which are complementary to collider experiments. In this talk, I have discussed the role of high-energy and ultra-high energy cosmic neutrinos in this endeavor.

Neutrino astronomy has two primary advantages which make such experiments complementary to colliders. Firstly, neutrino-nucleon interactions occur naturally and frequently with energies well beyond those accessible in any planned or proposed collider experiment. Secondly, cosmic neutrinos travel over incredibly long baselines: kiloparsecs, megaparsecs and gigaparsecs. These two characteristics allow cosmic neutrinos to be used to test aspects of fundamental physics which are difficult or impossible to explore with collider technology.

In this talk, I have considered a handful of particle physics models in which neutrino astronomy is useful. Scenarios with low-scale (TeV) gravity or with electroweak instanton induced interactions each result in very large deviations from the Standard Model charged plus neutral current prediction for the neutrino-nucleon scattering cross section at high-energies. Alternatively, scenarios involing decaying neutrinos or quantum decohernce can only be explored using measurements over extremely long baselines. The next generation of neutrino astronomy may very well provide a new window into phenomenon such as these.

\bigskip
{\small This work was supported by the US Department of Energy and by NASA grant NAG5-10842.}


\bigskip

\bbib{9}               



\bibitem{[1]} 
A.~Kusenko and T.~J.~Weiler,
Phys.\ Rev.\ Lett.\  {\bf 88}, 161101 (2002)
[arXiv:hep-ph/0106071].


\bibitem{[2]}
D.~Hooper,
Phys.\ Rev.\ D {\bf 65}, 097303 (2002)
[arXiv:hep-ph/0203239].

\bibitem{[3]}
R.~C.~Myers and M.~J.~Perry,
  Annals Phys.\  {\bf 172}, 304 (1986);
 P.~C.~Argyres, S.~Dimopoulos and J.~March-Russell,
  Phys.\ Lett.\ B {\bf 441}, 96 (1998)
  [arXiv:hep-th/9808138].


\bibitem{[4]}
P.~D.~D'Eath and P.~N.~Payne, 
Phys.\ Rev.\ D {\bf 46}, 658 (1992); 
Phys.\ Rev.\ D {\bf 46}, 675 (1992);
Phys.\ Rev.\ D {\bf 46}, 694 (1992). 


\bibitem{[5]}
  L.~A.~Anchordoqui, J.~L.~Feng, H.~Goldberg and A.~D.~Shapere,
  Phys.\ Rev.\ D {\bf 68}, 104025 (2003)
  [arXiv:hep-ph/0307228].

\bibitem{[6]}
 L.~Anchordoqui, M.~T.~Dova, A.~Mariazzi, T.~McCauley, T.~Paul, S.~Reucroft and J.~Swain,
  Annals Phys.\  {\bf 314}, 145 (2004)
  [arXiv:hep-ph/0407020].


\bibitem{[7]}
 R.~Engel, D.~Seckel and T.~Stanev,
  Phys.\ Rev.\ D {\bf 64}, 093010 (2001)
  [arXiv:astro-ph/0101216].


\bibitem{[8]}
L.~A.~Anchordoqui, J.~L.~Feng, H.~Goldberg and A.~D.~Shapere,
  Phys.\ Rev.\ D {\bf 65}, 124027 (2002)
  [arXiv:hep-ph/0112247].


\bibitem{[9]}
  L.~Anchordoqui, T.~Han, D.~Hooper and S.~Sarkar,
  arXiv:hep-ph/0508312.

\bibitem{[10]}
  J.~Alvarez-Muniz, J.~L.~Feng, F.~Halzen, T.~Han and D.~Hooper,
  Phys.\ Rev.\ D {\bf 65}, 124015 (2002)
  [arXiv:hep-ph/0202081];
  M.~Kowalski, A.~Ringwald and H.~Tu,
  Phys.\ Lett.\ B {\bf 529}, 1 (2002)
  [arXiv:hep-ph/0201139].


\bibitem{[11]}
  M.~Ahlers, A.~Ringwald and H.~Tu,
  arXiv:astro-ph/0506698.


\bibitem{[12]}
A.~Ringwald, 
Phys.\ Lett.\ B {\bf 555}, 227 (2003);
F.~Bezrukov, D.~Levkov, C.~Rebbi, V.~A.~Rubakov and P.~Tinyakov
Phys.\ Rev.\ D {\bf 68}, 036005 (2003);
F.~Bezrukov, D.~Levkov, C.~Rebbi, V.~A.~Rubakov and P.~Tinyakov,
Phys.\ Lett.\ B {\bf 574}, 75 (2003);
A.~Ringwald,
JHEP {\bf 0310}, 008 (2003). 


\bibitem{[13]}
  T.~Han and D.~Hooper,
  Phys.\ Lett.\ B {\bf 582}, 21 (2004)
  [arXiv:hep-ph/0307120].

\bibitem{[14]}
  T.~Kashti and E.~Waxman,
  arXiv:astro-ph/0507599.

\bibitem{[15]}
J.~N.~Bahcall, S.~T.~Petcov, S.~Toshev and J.~W.~F.~Valle,
Phys.\ Lett.\ B {\bf 181}, 369 (1986);
J.~F.~Beacom and N.~F.~Bell,
Phys.\ Rev.\ D {\bf 65}, 113009 (2002).


\bibitem{[16]}
J.~F.~Beacom, N.~F.~Bell, D.~Hooper, S.~Pakvasa and T.~J.~Weiler,
Phys.\ Rev.\ Lett.\  {\bf 90}, 181301 (2003).


\bibitem{[17]}
J.~F.~Beacom, N.~F.~Bell, D.~Hooper, S.~Pakvasa and T.~J.~Weiler,
Phys.\ Rev.\ D {\bf 68}, 093005 (2003).


\bibitem{[18]}
 D.~Hooper, D.~Morgan and E.~Winstanley,
  Phys.\ Lett.\ B {\bf 609}, 206 (2005)
  [arXiv:hep-ph/0410094].

\bibitem{[19]}
  N.~Hayashida {\it et al.}  [AGASA Collaboration],
  Astropart.\ Phys.\  {\bf 10}, 303 (1999)
  [arXiv:astro-ph/9807045].


\bibitem{[20]}
  L.~A.~Anchordoqui, H.~Goldberg, F.~Halzen and T.~J.~Weiler,
  Phys.\ Lett.\ B {\bf 593}, 42 (2004)
  [arXiv:astro-ph/0311002].



\bibitem{[21]}
 L.~A.~Anchordoqui, H.~Goldberg, M.~C.~Gonzalez-Garcia, F.~Halzen, D.~Hooper, S.~Sarkar and T.~J.~Weiler,
  arXiv:hep-ph/0506168.


\ebib
\end{document}